\documentclass[aps,pre,showkeys,amssymb]{revtex4-2}
\usepackage{graphicx}
\usepackage{color}
\usepackage[utf8]{inputenc}
\usepackage{braket}
\usepackage{graphicx}
\usepackage{appendix}

\begin{document}

\title{Fractional Schr\"odinger Equation and Time Dependent Potentials}

\author{E. C. Gabrick$^1,$\footnote{Corresponding Author \\e-mail: ecgabrick@gmail.com}, E. Sayari$^1$, A. S. M. de Castro$^{1,2}$, J. Trobia$^3$, A. M. Batista$^{1,3}$, E. K. Lenzi$^{1,2}$}
\affiliation{
$^{1}$Graduate Program in Science, State University of Ponta Grossa, Ponta Grossa, PR, Brazil\\
$^{2}$Department of Physics, State University of Ponta Grossa, Ponta Grossa, PR, Brazil\\
$^3$Department of Mathematics and Statistics, State University of Ponta Grossa, Ponta Grossa, PR, Brazil
}

\begin{abstract}
We investigate the solutions for a time-dependent potential by considering 
two scenarios for the fractional Schr\"odinger equation. The first scenario 
analyzes the influence of the time-dependent potential in the absence of 
the kinetic term. We obtain analytical and numerical solutions for this 
case by considering the Caputo fractional time derivative, which extends 
Rabi's model. In the second scenario, we incorporate the kinetic term in 
the Schr\"odinger equation and consider fractional spatial derivatives. 
For this case, we analyze the spreading of the Gaussian wave package 
under the action of the time and spatial fractional differential operators.
\end{abstract}
\keywords{anomalous spreading, fractional dynamics, fractional quantum mechanics}

\maketitle

\section{Introduction}
The meaning of the operator $d^\nu y /dx^\nu$ with $\nu$ integer is well 
known and has a profound physical background~\cite{bookguo}. The challenge 
is to understand what this operator means when $\nu$ is any number (positive or
negative, real or complex) \cite{Book01} or even a function~\cite{perez2018}. 
This problem can be dated from a letter of L'H\^opital to Leibniz in 1695, 
where he asked him what the operator   $d^\nu y /dx^\nu$ is when $\nu = 1/2$ \cite{Book01}. 
Since then, many researchers have dedicated themselves to this problem, 
for example, Euler, Lagrange, Laplace, Fourier, and others \cite{bookguo}, 
giving rise to the fractional calculus \cite{herrman2014}.

Nowadays, fractional calculus has quickly become a new efficient mathematical 
tool to analyze different properties of systems, in general, by extending 
the differential operators by incorporating non-integer indexes and, in 
particular, connecting them with experimental results~\cite{hernandez2021, ciuchi2012, bisquert2005, somer2020}.
In this manner, it is possible to investigate many situations with a 
simple extension that may incorporate memory effects, long-range correlations, 
and other effects in complex systems~\cite{ali2020}. For instance, in 
complex viscoelastic media~\cite{pandey2016physical,bagley1983theoretical}, 
electrical spectroscopy impedance~\cite{rosseto2022frequency,scarfone2022anomalous,lenzi2021anomalous}, 
wave propagation in porous media~\cite{chen2016causal,cai2018survey}, microflows 
of viscoelastic fluids~\cite{jiang2017transient}, and gas transport in 
heterogeneous media~\cite{chang2019spatial,chang2018time,pandey2016connecting}. 
It has also been used in other physics branches to extend several partial 
differential equations to cover and bring new possibilities for applications 
in different scenarios \cite{bookguo}. One of them is quantum mechanics, 
which has been extended by incorporating spatial and time fractional 
differential operators \cite{wang2007, heydari2023}. In this context, we 
have the pioneers works of N. Laskin~\cite{laskin2018fractional,PhysRevE.66.056108, laskin2000}, 
which lead us to a fractional Schr\"odinger equation, and that has been 
followed by other extensions incorporating fractional differential operators 
in time, and space~\cite{Book01} as well as non-local terms~\cite{doi10.10631.4894059,doi10.10631.2842069} 
and constraints among the different spatial coordinates 
(comb-model)~\cite{doi10.10631.4996573,doi10.10631.5079226,SANDEV20191695}. 
These extensions of the Schr\"odinger equation have also been analyzed by 
considering different choices of potential, such as delta potentials~\cite{oliveira2011}, 
constant or linear potentials~\cite{guo2006} and for some time dependent 
potentials~\cite{dong2014}.  It is worth mentioning that, from the analytical 
and numerical point of view, it is a challenge to obtain solutions when 
fractional time derivatives are considered. 

Our goal in this work is to investigate the implications of 
considering time dependent potentials in the following fractional 
Schr\"odinger equation~\cite{doi:10.1063/1.1769611}
\begin{eqnarray}
\label{FSEQ1}
i^{\alpha} \hbar_{\alpha} \frac{\partial^{\alpha}}{\partial t^{\alpha}}\psi (\vec{r},t) &=& \widehat{H}(t) \psi(\vec{r},t)\;,
\end{eqnarray}
where the fractional differential operator is the Caputo fractional time 
derivative, defined as follows~\cite{Book01}:
\begin{eqnarray}
\frac{\partial^{\alpha}}{\partial t^{\alpha}}\psi (\vec{r},t)=
\frac{1}{\Gamma\left(1-\alpha\right)}
\int_{0}^{t}dt'\frac{1}{(t-t')^{\alpha}}\frac{\partial}{\partial t}\psi (\vec{r},t),
\end{eqnarray}
for $0<\alpha<1$. We employ analytical and numerical approaches to analyze 
Eq.~(\ref{FSEQ1}). For the last one, we consider the ﬁnite difference 
method~\cite{murio2008, liu2005, rydin2021}. It should be mentioned, as 
discussed in Ref.~\cite{doi:10.1063/1.1769611}, that we can also extend 
the Schr\"odinger equation as follows:
\begin{eqnarray}
\label{FSEQ2a}
i\hbar_{\alpha} \frac{\partial^{\alpha}}{\partial t^{\alpha}}\psi (\vec{r},t) &=& \widehat{H}(t) \psi(\vec{r},t).
\end{eqnarray}
Equations~(\ref{FSEQ1}) and~(\ref{FSEQ2a}) are two possible extensions of 
the Schr\"odinger equation. However, when performing a Wick rotation, the 
imaginary unit is raised to the same power as the time coordinates for 
Eq.~(\ref{FSEQ1}). Another point between the two equations involves the 
temporal behavior of the solution, which for the first case, is more 
suitable than the second one that decreases or grows with time instead 
of a sinusoidal behavior. For these reasons point out in Ref.~\cite{doi:10.1063/1.1769611}, 
we consider Eq.~(\ref{FSEQ1}) in our developments. 
It is also interesting to mention the similar appearance between the 
Schr\"odinger and diffusion equations. This similarity between these 
equations is a consequence of the stochastic processes behind these equations, 
which can be evidenced by Feynman’s path integral formulation~\cite{FeynmanHibbs}, 
and transformed into Wiener’s path integral, which is the integral over 
the path of Brownian motions. It has also motivated different extensions 
motivated by other aspects, which include L\'evy distributions~\cite{laskin2000}, 
comb-model~\cite{Iomin2011,lenzi2022schrodinger}, among others. In addition, 
these extensions of the Schr\"odinger equations have been considered in 
problems related to optica~\cite{Okposo2022}, solutions for free-particle~\cite{Achar2013}, 
optical solitons~\cite{Esen2018} and others~\cite{Liaqat2022, Hilfer2000, Heydari2019, El-Nabulsi2023, Ain2020}.

By using Eq.~(\ref{FSEQ1}), we consider a two-level system with a time 
dependence on the potential and restricted to a one-dimensional wave 
function $\psi(x,t)$, without any loss of generality, and $\hbar_{\alpha}$ 
is an arbitrary time constant used to replace the Planck constant 
(see Ref.~\cite{doi:10.1063/1.1769611} for more
details). As mentioned before, the difference between the definitions given 
by Eq.~(\ref{FSEQ1}) and Eq.~(\ref{FSEQ2a}) is in the imaginary unit. 
Both equations violate the probability conservation law~\cite{zu2022}. 
However, the probability related to Eq.~(\ref{FSEQ1}) may increase and 
reach a constant value  $1/\alpha^{2}$ as discussed in Ref.~\cite{lu2017} 
and the probability associated with Eq.~(\ref{FSEQ2a}) decays to zero \cite{zu2022}.
It is worth mentioning that the two-level systems 
are very interesting because the simplicity and richness of 
results~\cite{sakurai} have been used to study spin 1/2-like~\cite{cohen},  
magnetic resonance~\cite{Ruyten1990},
quantum computation~\cite{Angelo2005}, unitary evolution~\cite{Cius2002}, 
and others~\cite{Itano1993}. In some cases, the two-level
systems are analytical soluble, mostly when the Hamiltonian is unperturbed. 
However, perturbed
Hamiltonians are particularly interesting, mainly in the presence of an 
electromagnetic field~\cite{Kibis2009}. In situations like that, i.e., 
the time-dependent Hamiltonian, the exact solution is rare; one famous 
example is the Rabi problem~\cite{Rabi1937}. Inspired by the Rabi problem 
and electromagnetic fields perturbation, we consider two distinct cases for the Hamiltonian in
Eq.~(\ref{FSEQ1}). The first one considers
\begin{equation}
\widehat{H}(t) = \left(\begin{array}{cc}
E_1 & \gamma e^{i\omega t}\\
\gamma e^{-i \omega t} & E_2
\end{array}\right),
\label{hamiltonian2d1}
\end{equation}
which corresponds to a two-level system, where $E_{1}$ and $E_{2}$ are 
the eigenvalues, and $\gamma$ is the amplitude of the external field 
with frequency equals $\omega$. In the second case, we consider the Hamiltonian given by 
\begin{equation}
\widehat{H}(t) = \left(\begin{array}{cc}
\widehat{p}^{\,\mu}/(2m) & \gamma e^{i\omega t}\\
\gamma e^{-i \omega t} & \widehat{p}^{\,\mu}/(2m)
\end{array}\right),
\label{hamiltonian2d11}
\end{equation}
which incorporates a kinetic term and consequently a spatial 
dependence in our problem. Note that the kinetic terms have the power $\mu$, 
which can be related to a spatial fractional derivative, i.e., 
${\cal{F}}^{-1}\{|p|^{\mu}\widetilde{\psi}(p,t)\}\equiv -\partial^{\mu}_{|x|}\psi(x,t)$, where ${\cal{F}}\{\psi(x,t);k\}=\widetilde{\psi}(k,t)$ (and ${\cal{F}}^{-1}\{\tilde{\psi}(k,t);x\}=\psi(x,t)$) 
corresponds to the Fourier transform, respectively. This definition corresponds 
to the Riesz derivative~\cite{evangelista2023introduction,Bayin2016}.

Aiming to understand the influence of fractional order in Schr\"odinger 
equation, we made the developments for standard quantum mechanics in 
Sec. II and for the fractional operators in Sec. III. We obtain analytical 
and numerical solutions for these Hamiltonians and analyze the spreading 
behavior of the wave package in different conditions. Finally, we present 
our discussions and conclusions in Sec. IV.


\section{Schr\"odinger Equation}

The standard Schr\"odinger equation is an specific case of 
Eq.~(\ref{FSEQ1}) or Eq.~(\ref{FSEQ2a}) with $\alpha = 1$. To understand 
the effects of $\alpha\neq 1$ in quantum dynamics, we first analyze the 
results obtained from the standard case. In this sense,
let us start our analysis by reviewing the results obtained for the 
standard Schr\"odinger equation, i.e.,
\begin{equation}
i \hbar \frac{\partial}{\partial t} \psi (\vec{r},t)  =  \widehat{H} \psi (\vec{r},t) \;,
    \label{schrodinger-eq}
\end{equation}
where $\widehat{H}$ is the Hamiltonian operator, $\psi (\vec{r},t)$ is 
the wave function, $i$ is the imaginary unit, and $\hbar$ is the Planck 
constant~\cite{sakurai}, which, for simplicity, we consider $\hbar = 1$. 
Equation~(\ref{schrodinger-eq}) will be analyzed first by considering the 
Hamiltonian given by Eq.~(\ref{hamiltonian2d1})
which corresponds to a two-level system, as previously discussed. 
Equation~(\ref{hamiltonian2d1}) has been applied in several situations, 
such as a two-level system interacting with light field~\cite{lu2017}. 
After, we incorporate kinetic terms in Eq.~(\ref{hamiltonian2d1}) by 
performing the following change $E_{1}\rightarrow \widehat{p}^{\,2}/\left(2m\right)$ and $E_{2}\rightarrow \widehat{p}^{\,2}/\left(2m\right)$, 
which implies
\begin{equation}
\widehat{H} = \left(\begin{array}{cc}
\widehat{p}^{\,2}/\left(2m\right) & \gamma e^{i\omega t}\\
\gamma e^{-i \omega t} & \widehat{p}^{\,2}/\left(2m\right)
\end{array}\right)\;.
\label{hamiltonian2d2}
\end{equation}
Equation~(\ref{hamiltonian2d2}) is equivalent to considering 
the particular case $\mu=2$ in Eq.~(\ref{hamiltonian2d11}), i.e., it 
considers the kinetics terms with an integer index.
After analyzing the standard Schr\"odinger equation which emerges from 
these cases, we consider the fractional extensions of these cases and 
analyze the implications for spreading the wave package, particularly the case $\mu\neq 2$.
Equations~(\ref{hamiltonian2d1}) and~(\ref{hamiltonian2d2}) allows us 
to consider that the wave function has the following form
\begin{eqnarray}
\psi=\left(\begin{array}{cc}
\psi_{1}\\
\psi_{2}\end{array}
\right)\;,
\end{eqnarray}
with $\psi_{1}$ and $\psi_{2}$ are obtained by solving the Schr\"odinger 
equation for each case.

Now let us consider the first case correspondent to the Hamiltonian, 
defined in terms of Eq.~(\ref{hamiltonian2d1}) and solutions 
$\psi_{k}=\psi_{k}(t)$. For the initial condition, we analyze the situation 
in which only one state is populated initially, i.e., the initial condition 
is given by $\psi_{1}(0)=1$ and $\psi_{2}(0)=0$. The problem concerns 
obtaining the probability transition between states after applying the 
external field. We find these probabilities by solving Eq.~(\ref{schrodinger-eq}), i.e.,
\begin{eqnarray}
\label{standard1a}
i \frac{\partial}{\partial t} \psi_1(t) &=& E_1 \psi_1(t) + \gamma e^{i \omega t} \psi_2(t),
\end{eqnarray}
and
\begin{eqnarray}
i \frac{\partial}{\partial t}  \psi_2(t) &=& E_2 \psi_2(t) + \gamma e^{-i \omega t} \psi_1(t),
\label{standard1b}
\end{eqnarray}
in which $|\psi_1 (t)|^2 + |\psi_2 (t)|^2 = 1$ is always verified. This 
case admits an analytical solution; for example, see Ref.~\cite{sakurai} 
and, in particular, when $\omega = 0$, it is given by
\begin{eqnarray}
\psi_{1} (t) = \frac{1}{2}\left(1 + \frac{E_1 - E_2}{\sqrt{(E_1 - E_2)^2 + 4\gamma^2}} \right)e^{-i\gamma_{+}t} + \frac{1}{2}\left(1 - \frac{E_1 - E_2}{\sqrt{(E_1 - E_2)^2 + 4\gamma^2}} \right)e^{-i\gamma_{-}t},
\end{eqnarray}
and
\begin{eqnarray}
\psi_{2} (t) = \frac{\gamma}{\sqrt{(E_1 - E_2)^2 + 4\gamma^2}}\left(e^{-i\gamma_{+}t} - e^{-i\gamma_{-}t}\right),
\end{eqnarray}
where $\gamma_{\pm}=\left(E_1 + E_2 \pm \sqrt{(E_1 - E_2)^2 + 4\gamma^2}\right)/2$. 

\begin{figure}[hbt]
\centering
\includegraphics[scale=0.4]{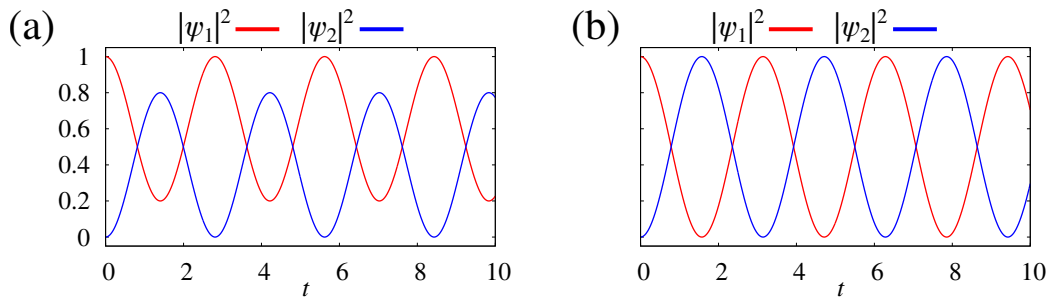}
\caption{Probability of finding the system in $\psi_1 (t)$ state given 
by the red line and in the $\psi_2(t)$ by the blue line. The panel (a) 
is for $\omega = 0$ and (b) for $\omega = 1$. We consider $\gamma = 1$, 
$E_1 = 1$ and $E_2 = 2$.}
\label{fig1}
\end{figure}

Figure~\ref{fig1}(a) illustrates the result obtained by considering the 
external field constant, i.e., $\omega=0$. It is possible to verify that 
the system has oscillation between two levels. 
It is interesting to note in Fig.~\ref{fig1}(a) that for the parameters 
used, $\psi_1$ is predominant over $\psi_{2}$. On the other hand, when 
we consider an oscillatory external field, $\omega \neq 0$, the system 
oscillates between the two states as shown in Fig.~\ref{fig1}(b). The 
result for $\omega \neq 0$ is numerically obtained by solving Eqs.~(\ref{standard1a}) 
and~(\ref{standard1b}). When $0<\omega<1$, the amplitude of $|\psi_2|^2$ tends 
to 1, founding this value in $\omega = 1$. By the other hand, for $\omega > 1$ 
the $|\psi_2|^2$ value oscillate asymptotically to zero while $|\psi_1|^2$ 
oscillate in 1 direction. 

Similar analysis can be performed by considering the second case, i.e., 
for the Hamiltonian defined in terms of  Eq.~(\ref{hamiltonian2d2}). The 
equations for the wave functions 
$\psi_{1(2)}(x,t)$ read as
\begin{eqnarray}
\label{standard2a}
i \frac{\partial}{\partial t} \psi_1(x,t) &=& -\frac{1}{2} \frac{\partial^{2}}{\partial x^{2}} \psi_1(x,t) + \gamma e^{i \omega t} \psi_2(x,t),
\end{eqnarray}
and
\begin{eqnarray}
i \frac{\partial}{\partial t}  \psi_2(x,t) &=& -\frac{1}{2} \frac{\partial^{2}}{\partial x^{2}} \psi_2(x,t) + \gamma e^{-i \omega t} \psi_1(x,t),
\label{standard2b}
\end{eqnarray}
where, for simplicity, we assume $m = 1$. The solution for this case can 
be found by applying the Fourier transform ($
\widetilde{\psi}_{1,2}(k,t)={\cal{F}}\{\psi_{1,2}(x,t);k\}$ and $\psi_{1,2}(x,t)
={\cal{F}}^{-1}\{\widetilde{\psi}_{1,2}(k,t);x\}$), as defined before) in
Eqs.~(\ref{standard2a}) and~(\ref{standard2b}) yielding
\begin{eqnarray}
\label{standard2aa}
i \frac{\partial}{\partial t} \widetilde{\psi}_1(k,t) &=& \frac{1}{2}k^{2}  \widetilde{\psi}_1(k,t) + \gamma e^{i \omega t} \widetilde{\psi}_2(k,t), 
\end{eqnarray}
and
\begin{eqnarray}
i \frac{\partial}{\partial t}  \widetilde{\psi}_2(k,t) &=& \frac{1}{2} k^{2} \widetilde{\psi}_2(k,t) + \gamma e^{-i \omega t} \widetilde{\psi}_1(k,t)\;.
\label{standard2bb}
\end{eqnarray}
By performing some calculations, it is possible to show that the solution  
$\widetilde{\psi}_{2}(k,t)$ is related to the solution  $\widetilde{\psi}_{1}(k,t)$ as follows:
\begin{eqnarray}
\widetilde{\psi}_2(k,t)=-i\gamma\int_{0}^{t}dt'e^{-\frac{1}{2}ik^{2}(t-t')}e^{-i\omega t'}\widetilde{\psi}_{1}(k,t')\;,
\label{standard2bbb}
\end{eqnarray}
for which we assume the initial condition $\widetilde{\psi}_{2}(k,0)=0$. 
Furthermore, this relation implies that 
\begin{eqnarray}
\label{standard2aaa}
i \frac{\partial}{\partial t} \widetilde{\psi}_1(k,t) = \frac{1}{2}k^{2}  \widetilde{\psi}_1(k,t) - i\gamma^{2}\int_{0}^{t}dt'e^{-\frac{1}{2}ik^{2}(t-t')}e^{-i\omega (t-t')}\widetilde{\psi}_{1}(k,t')\;.
\end{eqnarray}
Note that the last term present in  Eq.~(\ref{standard2aaa}) is a nonlocal 
term and the kernel has a nonsingular dependence on the variable 
$t$. It is worth mentioning that the nonsingular kernels have been successfully 
applied in many situations, such as the ones presented in 
Refs.~\cite{Vinales2007, Vinales2008, Fa2009, Camargo2009, Vinales2009, Camargo2009b}.

Equation~(\ref{standard2aaa}) can be solved by using the Laplace 
transform, yielding
\begin{eqnarray}
\widetilde{\psi}_{1}(k,t)=e^{-\frac{1}{2}i\left(k^{2}-\omega\right)t}\left[\cos\left(\frac{1}{2}t\sqrt{\omega^{2}+4\gamma^{2}}\right)-\frac{i\omega}{\sqrt{\omega^{2}+4\gamma^{2}}}\sin\left(\frac{1}{2}t\sqrt{\omega^{2}+4\gamma^{2}}\right)\right]\widetilde{\varphi}_{1}(k)\;,
\label{standard2aaaa}
\end{eqnarray}
where $\widetilde{\psi}_{1}(k,0)=\widetilde{\varphi}_{1}(k)$ is the initial 
condition for $\psi_{1}(x,t)$. Applying the inverse of the Fourier transform,
we obtain that
\begin{eqnarray}
\psi_{1}(x,t)=\Xi_{1}(t)\int_{-\infty}^{\infty}dx'{\cal{G}}(x-x',t)\varphi_{1}(x'),
\label{standard-analitica1}
\end{eqnarray}
where 
\begin{eqnarray}
\Xi_{1}(t)=e^{\frac{i}{2}\omega t} \left[\cos\left(\frac{1}{2}t\sqrt{\omega^{2}+4\gamma^{2}}\right)-\frac{i\omega}{\sqrt{\omega^{2}+4\gamma^{2}}}\sin\left(\frac{1}{2}t\sqrt{\omega^{2}+4\gamma^{2}}\right)\right],
\end{eqnarray}
and
\begin{eqnarray}
\psi_{2}(x,t)=\Xi_{2}(t)\int_{-\infty}^{\infty}dx'{\cal{G}}(x-x',t)\varphi_{1}(x')\;.
\label{standard-analitica2}
\end{eqnarray}
The function $\Xi_{2}(t)$ is written as follows: 
\begin{eqnarray}
\Xi_{2}(t)=-\frac{2i\gamma }{\sqrt{\omega^{2}+4\gamma^{2}}}e^{-\frac{i}{2}\omega t}\sin\left(\frac{1}{2}t\sqrt{\omega^{2}+4\gamma^{2}}\right),
\end{eqnarray}
and ${\cal{G}}(x,t)$ is the quantum free particle propagator, i.e., 
${\cal{G}}(x,t)=e^{-\frac{x^{2}}{2it}}/\sqrt{2\pi it}$. 

In addition to the analytical result, given by the Eqs.~(\ref{standard-analitica1}) 
and (\ref{standard-analitica2}), we also obtain the numerical solutions of the 
Eqs.~(\ref{standard2a}) and (\ref{standard2b}). For the numerical approach, 
we consider the finite difference method \cite{crank1975}. We consider a 
grid defined by $[0,X] \times [0,T]$, with boundary conditions 
equal to $\psi_{1,2}(0,t)=\psi_{1,2}(X,t)=0$. The time is discretized by 
$t_j = j\Delta t$, where $j=1,2,...,N_t$, with time step equal to $\Delta t = T/N_t$; 
and the each space coordinate is given by $x_i = i\Delta x$, where $i = 1,2,...,N_x$, 
with space step equal to $\Delta x = X/N_x$. To avoid numerical boundary problems, 
the origin of our space coordinate is in $X/2$. From these considerations, the 
discretization of Eqs.~(\ref{standard2a}) and (\ref{standard2b}) are given by
\begin{equation}
 \label{standardnumerical-a}
\psi^{i,j+1}_1 = \psi^{i,j}_1 + i\xi(\psi^{i+1,j}_1 - 2\psi^{i,j}_1 + \psi^{i-1,j}_1) - i\beta(V^{j}_1 \psi^{i,j}_2 + V^{j+1}_1 \psi^{i,j+1}_2),   
\end{equation}
and
\begin{equation}
\psi^{i,j+1}_2 = \psi^{i,j}_2 + i\xi(\psi^{i+1,j}_2 - 2\psi^{i,j}_2 + \psi^{i-1,j}_2) - i\beta(V^{j}_2 \psi^{i,j}_1 + V^{j+1}_2 \psi^{i,j+1}_1),
\label{standardnumerical-b}
\end{equation}
where $\xi\equiv \Delta t /2\Delta x^2$, $\beta \equiv \gamma\Delta t / 2$, $V_1 = e^{i\omega t}$ and $V_2 = e^{-i \omega t}$. For the stability conditions, it is required that $\xi\leq 1/2$ and the $\beta$ order less than $\xi$ order \cite{crank1975}.  

\begin{figure}[hbt]
\centering
\includegraphics[scale=0.28]{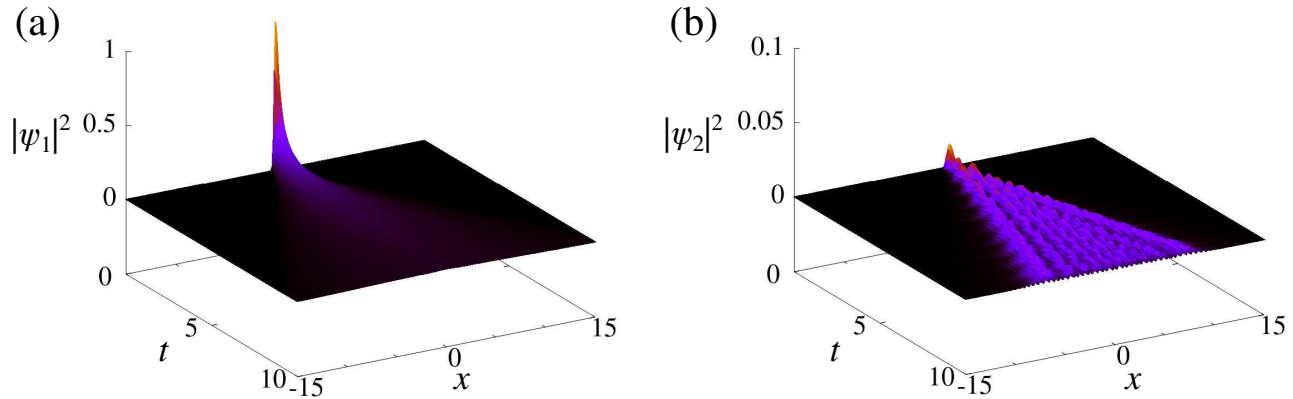}
\caption{Probability distribution with a kinetic term of finding the system 
in $\psi_1$ state, in the panel (a), and in $\psi_2$ state, in the panel (b). 
The initial condition is given by  $\psi_1(x,0) = e^{-\frac{x^2}{2\sigma^2}}/(2\pi\sigma^2)^{1/4}$ and $\psi_2(x,0) = 0$. 
We consider $\xi = 0.0016$, $\gamma=0.5$, $\sigma=0.4$, $\omega=2\pi$, $\Delta x = 0.25$, $\Delta t = 0.0002$, and $\sigma = 0.4$.}
\label{fig2}
\end{figure}

Considering  $\psi_1(x,0) = e^{-\frac{x^2}{2\sigma^2}}/(2\pi\sigma^2)^{1/4}$ 
and $\psi_2(x,0) = 0$ as the initial condition, the results for $|\psi_1|^2$ 
and $|\psi_2|^2$ are displayed in Figs.~\ref{fig2}(a) and (b), respectively. 
The parameters considered in this simulation are: $\xi = 0.0016$, 
$\gamma = 0.5$, $\Delta x = 0.25$, $\Delta t = 0.0002$, and $\sigma = 0.4$. 
As observed in the results without kinetic terms, the system starts mostly 
in $\psi_1$. However, a transition occurs to $\psi_2$ state due to the external 
field. This effect is present in the presence of kinetic terms. Numerically, 
we observed that $\int_{-\infty}^{\infty}dx(|\psi_1 (x,t)|^2 + |\psi_2 (x,t)|^2) = 1$. 
 It is worth mentioning that if we decrease $\Delta x$, the oscillations 
due to the potential become smoother.

\begin{figure}[hbt]
\centering
\includegraphics[scale=0.28]{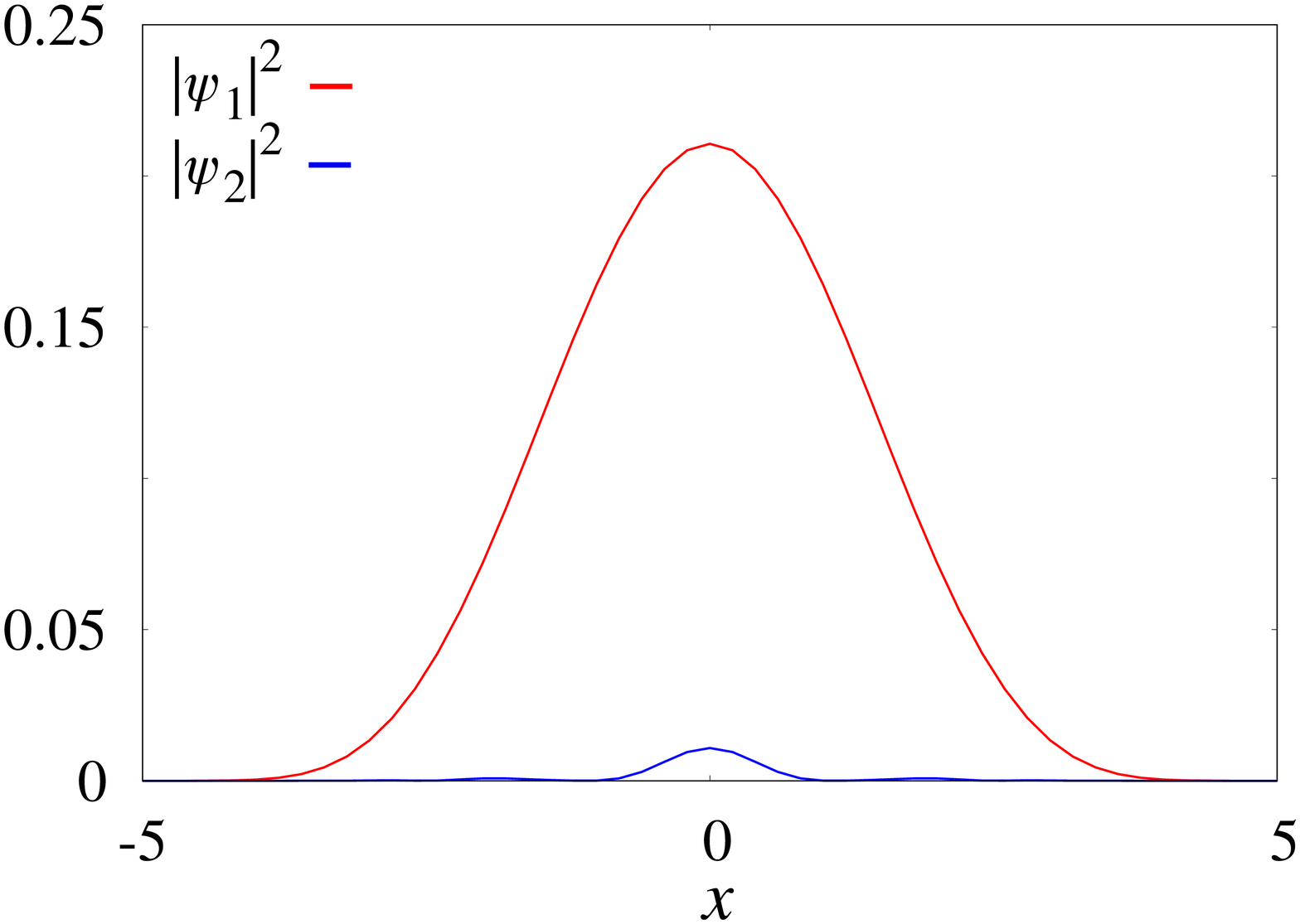}
\caption{Probability distribution at $t = 1.5$, for the states $\psi_1$ 
(red line) and $\psi_2$ (blue line). We consider $\xi = 0.0012$, 
$\gamma=0.5$, $\sigma=0.4$, $\omega=2\pi$, $\Delta x = 0.2$, $\Delta t = 0.0001$, and $\sigma = 0.4$.}
\label{fig3}
\end{figure}

The probability of finding the system in both states becomes approximately 
equal after $t \geq 10$. The first state is mostly populated for a short 
time, as observed in the result in Fig.~\ref{fig3}. This result shows 
that the package centered in origin is spread in the space in the first 
state and starts transit to the second state in a sinusoidal form.

The mean square displacement is a measure of the spreading 
of the system, represented by the wave package. It is widely applied in 
diffusion processes to characterize the type of diffusion, usual or anomalous. 
For the usual diffusion, we have a linear time dependence for the mean square 
displacement, i.e., $\langle(\Delta x)^2\rangle\sim t$, which is related 
to the Markovian processes. For the anomalous case, we have that 
$\langle(\Delta x)^2\rangle \sim t^{S_d}$, where $S_d > 1$ and $S_d<1$ are 
related to the super-diffusive and sub-diffusive cases~\cite{Book01}, 
respectively. In quantum mechanics, we can also use this quantity to 
understand the spreading of the probability density, i.e., $|\psi_{1,2}|^2$, 
in time. The normal case corresponds to the free particle for the standard 
Schr\"odinger equation, where $\langle(\Delta x)^2\rangle\sim t^S$ with $S=2$. 
The anomalous cases are those that have different behaviors for the mean 
square displacement. Note that these results are in agreement with the 
analytical results obtained for Eq.~(\ref{standard2a}) and~(\ref{standard2b}), 
which results in Gaussian distributions for both wave functions. 

\begin{figure}[hbt]
\centering
\includegraphics[scale=0.28]{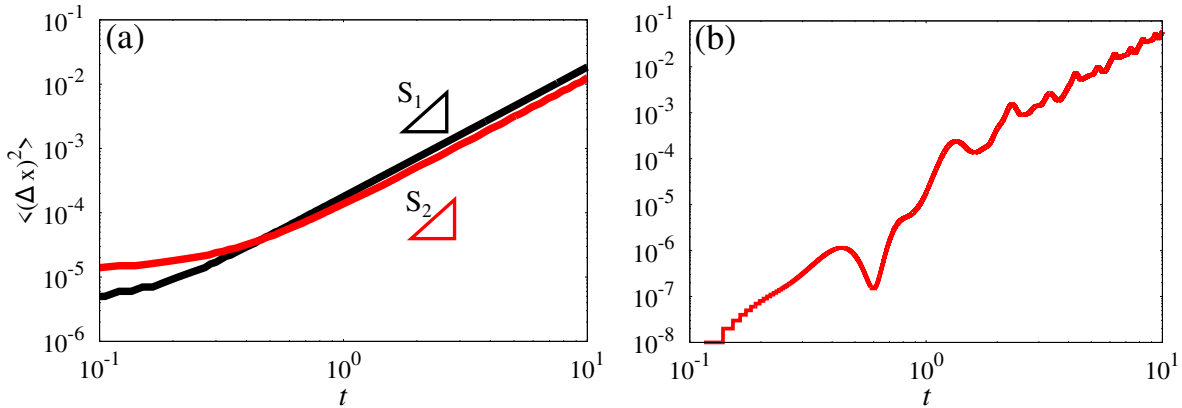}
\caption{Mean square displacement for the Gaussian package. Panel (a) 
is for $\psi_1$ state, and panel (b) is for $\psi_2$ standard case. 
The red points are for standard two-level equations. The black points 
are for the free particle. The slopes are $S_1 = 2.02$, $S_2 = 2.07$. 
We consider $\xi = 0.0012$, $\gamma=0.5$, $\sigma=0.4$, $\omega=2\pi$, $\Delta x = 0.2$, $\Delta t = 0.0001$, and $\sigma = 0.4$.}
\label{fig4}
\end{figure}

Considering the Gaussian package as the initial condition, 
the mean square displacement for the free particle is shown in Fig.~\ref{fig4}(a) 
by the black points, which follows $\sim t^{S_1}$, with $S_1 = 2.02$. This 
result is obtained by taking $\gamma = 0$ in the numerical simulations.
The effect of the  potential is displayed in Fig.~\ref{fig4}(a) by the 
red points, which follows $\sim t^{S_2}$ with $S_2 = 2.07$, for $|\psi_1|^2$.  
Due to the external potential, after a certain time, the probability of 
finding the system transfer from the first level to the second one, as 
shown in Fig.~\ref{fig4}(b). The distribution for $|\psi_2|^2$ increase 
as $\sim 2$. The slopes found by the numerical simulations are in agreement 
with our analytical expressions, which indicate $\sim t^{2}$ for both 
cases, free-particle and two-level system.


\section{Fractional Schr\"odinger Equations}

\begin{figure}[hbt]
\centering
\includegraphics[scale=0.35]{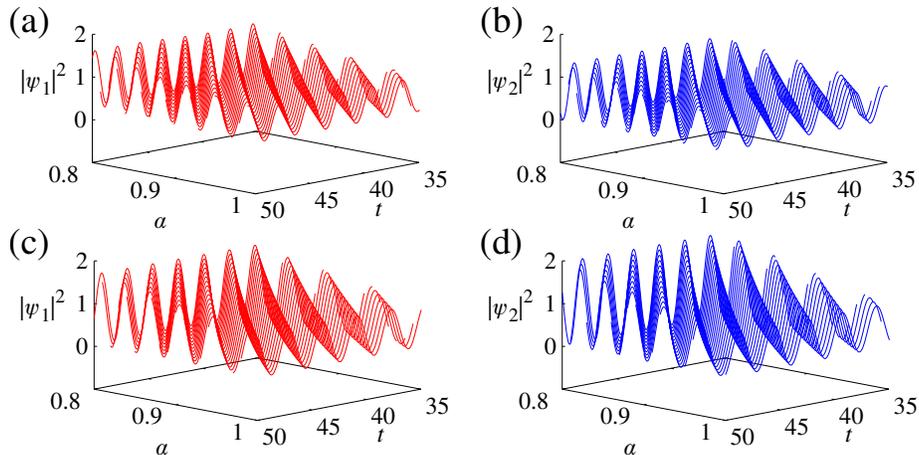}
\caption{Probability of finding the system in $\psi_1 (t)$ state given 
by the red line, in panels (a) and (c), and in the $\psi_2(t)$ by the 
blue line, in panels (b) and (d). The panel (a) is for $\omega = 0$ and 
(b) for $\omega = 1$. We consider $E_1 = 1$, $E_2 = 2$, $\gamma = 1$.}
\label{frac-two}
\end{figure}

Now, we analyze the previous scenarios within the fractional extension in 
time of the Schr\"odinger equation. For the first case, i.e., the 
Hamiltonian given by Eq.~(\ref{hamiltonian2d1}), we have that
\begin{eqnarray}
\label{fractional-twolevela}
i^\alpha \frac{\partial^\alpha}{\partial t^\alpha} \psi_1(t) &=& E_1 \psi_1(t) + \gamma e^{i \omega t} \psi_2(t), 
\end{eqnarray}
and
\begin{eqnarray}
i^\alpha \frac{\partial^\alpha}{\partial t^\alpha} \psi_2(t) &=& E_2 \psi_2(t) + \gamma e^{-i \omega t} \psi_1(t),
\label{fractional-twolevelb}
\end{eqnarray}
where $\hbar_{\alpha}=1$, without loss of generality. The relations 
represented by Eqs.~(\ref{fractional-twolevela}) and~(\ref{fractional-twolevelb})  
extend Rabi’s model. For this set of equations, obtaining an analytical 
solution for $\omega=0$, i.e., a static field, is possible. To obtain 
the solutions for this case, we can use the Laplace transform 
(${\cal{L}}\{\psi(t);s\}=\hat{\psi}(s)$ and ${\cal{L}}^{-1}\{\hat{\psi}(s);t\}=\psi(t)$) to 
simplify Eq.~(\ref{fractional-twolevela}) and~(\ref{fractional-twolevelb}) for 
the static case, yielding
\begin{eqnarray}
\label{fractional-twolevelaa}
\hat{\psi}_1(s)= \frac{i^{\alpha} s^{\alpha-1}\left(i^{\alpha} s^\alpha -E_2\right)}{\left(i^{\alpha} s^\alpha -E_1\right)\left(i^{\alpha} s^\alpha -E_2\right)-\gamma^{2}},
\end{eqnarray}
and
\begin{eqnarray}
\hat{\psi}_2(s) =\frac{i^{\alpha} s^{\alpha-1}}{\left(i^{\alpha} s^\alpha -E_1\right)\left(i^{\alpha} s^\alpha -E_2\right)-\gamma^{2}},
\label{fractional-twolevelbb}
\end{eqnarray}
for the initial condition $\psi_1(0)=1$ and $\psi_{2}(0)=0$. After 
performing the inverse of the Laplace transform, it is possible to show that
\begin{eqnarray}
\label{sol1a}
\psi_{1} (t) = \frac{1}{2}\left(1 - \frac{E_1 - E_2}{\sqrt{(E_1 - E_2)^2 + 4\gamma^2}} \right)E_{\alpha}\left(\gamma_{-}t^{\alpha}/i^{\alpha}\right) +\frac{1}{2}\left(1 + \frac{E_1 - E_2}{\sqrt{(E_1 - E_2)^2 + 4\gamma^2}} \right)E_{\alpha}\left(\gamma_{+}t^{\alpha}/i^{\alpha}\right),
\end{eqnarray}
and
\begin{eqnarray}
\label{sol1b}
\psi_{2} (t) = \frac{\gamma}{\sqrt{(E_1 - E_2)^2 + 4\gamma^2}}\bigg(E_{\alpha}\left(\gamma_{+}t^{\alpha}/i^{\alpha}\right) - E_{\alpha}\left(\gamma_{-}t^{\alpha}/i^{\alpha}\right)\bigg)\;,
\end{eqnarray}
where $\gamma_{\pm}=\left(E_1 + E_2 \pm \sqrt{(E_1 - E_2)^2 + 4\gamma^2}\right)/2$ and $E_{\alpha}(x)$ is the Mittag-Leffler function,
\begin{eqnarray}
E_{\alpha}(x)=\sum_{n=0}^{\infty}\frac{x^{n}}{\Gamma(1+\alpha n)}\;,
\end{eqnarray}
which corresponds to an extension of the exponential function~\cite{Book01}. 
The solutions found for $\psi_1(t)$ and $\psi_{2}(t)$, are determined in 
terms of the Mittag-Leffler function, implying that the system has an 
unusual oscillation process, i.e., different from the standard case.
For the case $\omega\neq 0$, the solution can also be found, and it is given by
\begin{eqnarray}
\label{sol2a1}
\psi_{1}(t)&=&E_{\alpha}\bigg[\left(E_{1}/i^{\alpha}\right)t^{\alpha}\bigg]\nonumber \\ &+&
\sum_{n=1}^{\infty}\left(\frac{\gamma}{i^{\alpha}}\right)^{2n}\int_{0}^{t}dt_{n}\Lambda(t-t_{n})\int_{0}^{t_{n}}dt_{n-1}\Lambda(t_{n}-t_{n-1})\cdots \int_{0}^{t_{2
}}dt_{t_{1}}\Lambda(t_{2}-t_{1})E_{\alpha}\bigg[\left(E_{1}/i^{\alpha}\right)t_{1}^{\alpha}\bigg],
\end{eqnarray}
with
\begin{eqnarray}
\label{sol2b1}
\psi_{2}(t)=\frac{\gamma}{i^{\alpha}}\int_{0}^{t}dt't'^{\alpha-1}E_{\alpha,\alpha}\bigg[\left(E_{2}/i^{\alpha}\right)(t-t')^{\alpha}\bigg]e^{i\omega t'}\psi_1(t')\;,
\end{eqnarray}
where
\begin{eqnarray}
\Lambda(t)=e^{i\omega t}\int_{0}^{t}dt't'^{\alpha-1}e^{-i\omega t'}E_{\alpha,\alpha}\bigg[\left(E_{1}/i^{\alpha}\right)t'^{\alpha}\bigg](t-t')^{\alpha-1}E_{\alpha,\alpha}\bigg[\left(E_{2}/i^{\alpha}\right)(t-t')^{\alpha}\bigg]\;,
\end{eqnarray}
by considering $\psi_1 (0) = 1$ and $\psi_2 (0) = 0$. The solutions for 
this case are found in terms of the generalized Mittag-Leffler function
~\cite{Book01},
\begin{eqnarray}
E_{\alpha,\beta}(x)=\sum_{n=0}^{\infty}\frac{x^{n}}{\Gamma(\beta+\alpha n)}\;.
\end{eqnarray}
Figure~\ref{frac-two} displays the numerical solution of Eqs.~(\ref{fractional-twolevela}) 
and (\ref{fractional-twolevelb}). For the static case, in Figs.~\ref{frac-two}(a) 
and~\ref{frac-two}(b), and for the non-static case, in Figs.~\ref{frac-two}(c) 
and~\ref{frac-two}(d). The results are in perfect agreement with the analytical 
solutions found in 
Eqs.~(\ref{sol1a}),~(\ref{sol1b}),~(\ref{sol2a1}), and~(\ref{sol2b1}) 
(see the Appendix for details of the numerical procedure). A direct 
consequence obtained by incorporating  fractional time derivative in the 
Schr\"odinger equation is the non-conservation of the probability, i.e., $|\psi_1 (\infty)|^2 + |\psi_2 (\infty)|^2 = 1/\alpha^2$.
This result agrees with the results presented in Refs.~\cite{lu2017, zu2022}.

Considering the kinetic terms, the time fractional Schr\"odinger equation 
can be written in the form
\begin{equation}
 \label{frac2a}
i^\alpha \frac{\partial^\alpha}{\partial t^\alpha} \psi_1(x,t) = -\frac{1}{2} \frac{\partial^{2}}{\partial x^{2}} \psi_1(x,t) + \gamma e^{i \omega t} \psi_2(x,t),  
\end{equation}
and
\begin{equation}
i^\alpha \frac{\partial^\alpha}{\partial t^\alpha}  \psi_2(x,t) = -\frac{1}{2} \frac{\partial^{2}}{\partial x^{2}} \psi_2(x,t) + \gamma e^{-i \omega t} \psi_1(x,t).
\label{frac2b}
\end{equation}
These equations can be approximated by the following discretization \cite{murio2008}:
\begin{eqnarray}
\psi^{i,j+1}_1 &=& \psi^{i,j}_1 - i^{-\alpha}\xi_\alpha(\psi^{i+1,j}_1 - 2\psi^{i,j}_1 + \psi^{i-1,j}_1) + i^{-\alpha}\beta_\alpha(V^{j}_1 \psi^{i,j}_2 + V^{j+1}_1 \psi^{i,j+1}_2) \nonumber \\
 &-& \sum_{k=1}^{j} [(k+1)^{(1-\alpha)} - k^{(1-\alpha)}][\psi^{i,j+1-k}_{1} - \psi^{i,j-k}_1], \\
 \end{eqnarray}
 and
 \begin{eqnarray}
\psi^{i,j+1}_2 &=& \psi^{i,j}_2 - i^{-\alpha}\xi_\alpha(\psi^{i+1,j}_2 - 2\psi^{i,j}_2 + \psi^{i-1,j}_2) + i^{-\alpha}\beta_\alpha(V^{j}_2 \psi^{i,j}_1 + V^{j+1}_2 \psi^{i,j+1}_1) \nonumber\\
&-& \sum_{k=1}^{j} [(k+1)^{(1-\alpha)} - k^{(1-\alpha)}][\psi^{i,j+1-k}_{2} - \psi^{i,j-k}_2], 
\end{eqnarray}
where $\xi_\alpha \equiv \Gamma(2-\alpha)\Delta t^\alpha /2\Delta x^2$, $\beta_\alpha \equiv \Gamma(2-\alpha)\gamma\Delta t^\alpha / 2$, $V_1 = e^{i\omega t}$ and $V_2 = e^{-i \omega t}$. The convergence condition is $\Delta t^\alpha / \Delta x^2 \leq (1 - 2^{-\alpha})/\Gamma(2-\alpha)$ \cite{liu2005}. 

\begin{figure}[hbt]
\centering
\includegraphics[scale=0.28]{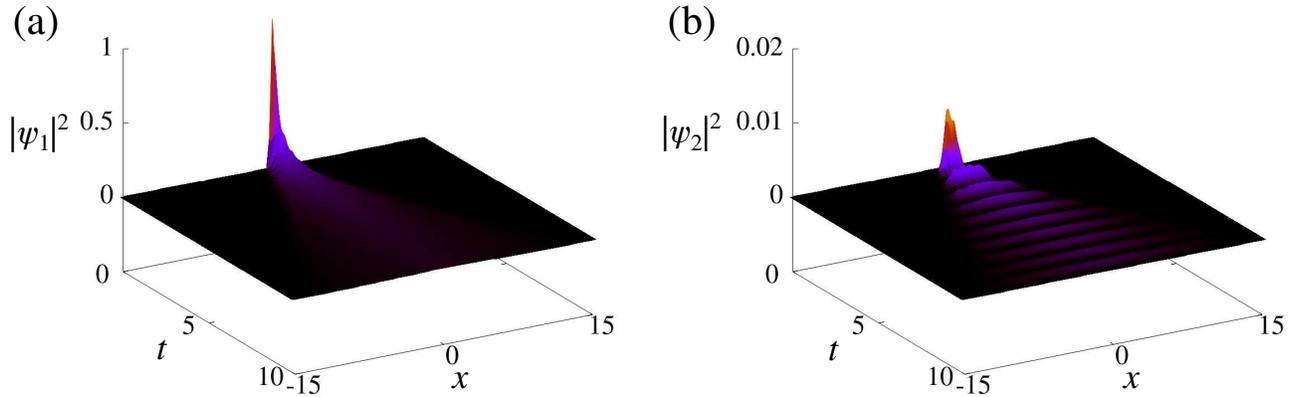}
\caption{Probability distribution with a kinetic term of finding the system 
in $\psi_1$ state, in the panel (a), and in $\psi_2$ state, in the panel (b), 
in the time fractional approach. We consider $\alpha = 0.98$, $\xi_\alpha = 0.0012$, 
$\beta_\alpha = 10^{-4}$, $\gamma=0.5$, $\sigma=0.4$, $\omega=2\pi$, $\Delta x = 0.5$, $\Delta t = 0.0005$.}
\label{fig6}
\end{figure}

Figures~\ref{fig6}(a) and~\ref{fig6}(b) show the numerical 
solution for $\psi_1(x,t)$ and $\psi_2(x,t)$ with $\alpha = 0.98$, by 
considering the initial conditions $\psi_1(x,0) = e^{-x^2/(2\sigma^2)}/(2\pi\sigma^2)^{1/4}$ and $\psi_2(x,0) = 0$, 
where $\sigma = 0.4$. Note that, for $\alpha$ slightly different from the 
standard case the dynamics properties of probabilities densities spreads 
have a significant change from the standard case. If we consider $\alpha <0.98$ 
these changes will be pronounced. For example, the results presented in 
Fig.~\ref{fig6}(a) and \ref{fig6}(b) show that the time fractional operator 
makes the probability spread slowly when compared to the standard case. 
Also, the transition between the states occurs with a greater amplitude 
than in the integer case. Another anomalous behavior is non-probability 
conservation. In this case, the probability decays, and the imaginary part 
of the effective potential operates as a dissipate term~\cite{bayin2013}.

\begin{figure}[hbt]
\centering
\includegraphics[scale=0.28]{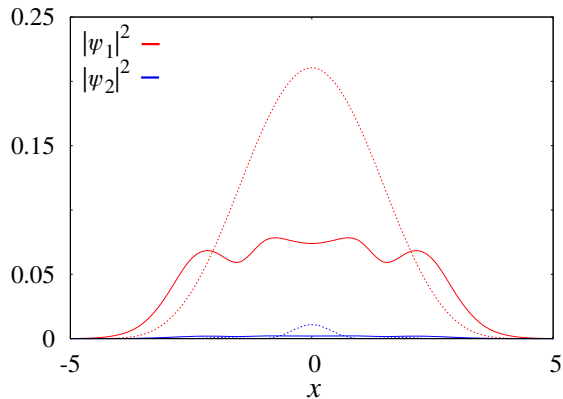}
\caption{Probability distribution at $t = 1.5$, for the states $\psi_1$ 
(red line) and $|\psi_2|^2$ (blue line). The continuous line shows the 
behavior for $\alpha = 0.98$ and the dotted for $\alpha = 1$. We consider 
$\alpha = 0.98$, $\xi_\alpha = 0.0012$, $\beta_\alpha = 10^{-4}$, $\gamma=0.5$, 
$\sigma=0.4$, $\omega=2\pi$, $\Delta x = 0.5$, $\Delta t = 0.0005$.}
\label{fig7}
\end{figure}

For a fixed time, a comparison between the probability distribution at 
the space in the case where $\alpha = 1$ (dotted lines) and $\alpha = 0.98$ 
(continuous lines) is shown in Fig.~\ref{fig7}. For this time, the results 
show that the package spread decreases the amplitude of $|\psi_1|^2$, 
and the shape of $|\psi_2|^2$ is wider.

\begin{figure}[hbt]
\centering
\includegraphics[scale=0.28]{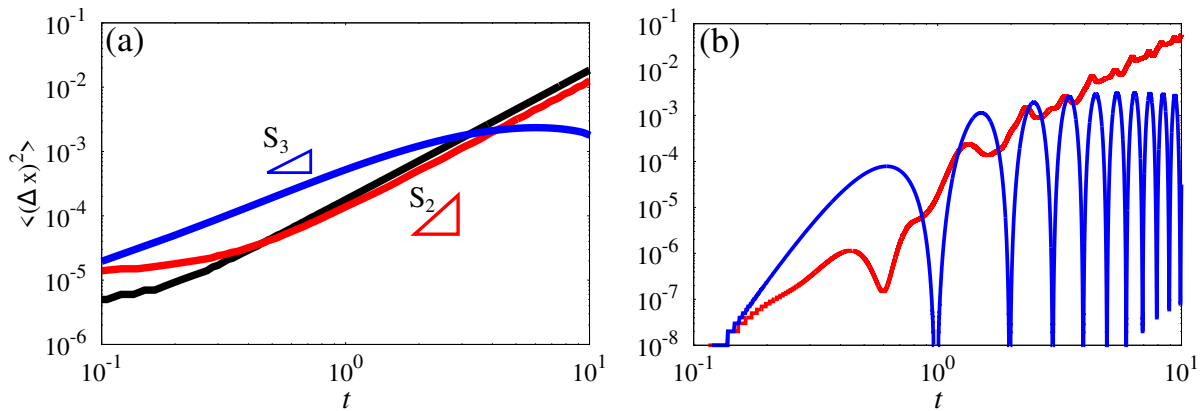}
\caption{Mean square displacement for the Gaussian package. Panel (a) 
is for $\psi_1$ state, and panel (b) is for $\psi_2$ state. The blue 
points are for $\alpha = 0.98$, the red for standard equations ($S_2 = 2.07$), 
and the black for a free particle. The slope associated with $\alpha = 0.98$ 
is $S_3 = 1.87$. We consider $\alpha = 0.98$, $\xi_\alpha = 0.0012$, 
$\beta_\alpha = 10^{-4}$, $\gamma=0.5$, $\sigma=0.4$, $\omega=2\pi$, $\Delta x = 0.5$, $\Delta t = 0.0005$.}
\label{fig8}
\end{figure}

The mean square displacement for the Gaussian package is exhibited in 
Fig.~\ref{fig8} for $\alpha = 0.98$ by the blue, in panel \ref{fig8}(a) 
for $|\psi_1|^2$ and in panel \ref{fig8}(b) for $|\psi_2|^2$. The red 
points are for the standard case, and the black line is for the free 
particle. The fractional time spread is similar to the standard 
case for short times. However, after this initial time, the blue points 
follow $\sim t^{S_3}$, with $S_3 = 1.87$, while the red ones $\sim t^{2.07}$. 
The behavior for the fractional case in time shows that the package spreads 
with less intensity than the standard case; the spread is more centered. 
The effect of the oscillatory potential is observed in the spread for the 
second state, as shown in Fig.~\ref{fig8}(b), by the blue curve for $|\psi_2|^2$. 
The second state populated in fractional presence in this scenario differs 
from the standard case. The fractional operator in time makes the probability 
for $\psi_2$ state oscillates like a sinusoidal function. As the time 
increase, the $\psi_2$ becomes populated with more frequency. Another 
difference in fractional cases is that the probability is non-conservative, 
and the deviations go to zero. The imaginary part of the effective potential 
operates like a dissipate term~\cite{bayin2013}.

Now, let us consider the  Schr\"odinger equation with fractional 
differential operators in space. This extension can be directly related to 
the works of  Laskin~\cite{laskin2000}, which takes L\'evy flights in the  
Feynman path integral approach into account. Following an analogous scheme~\cite{wang2007}, 
it is possible to include the fractional differential operator in space in 
such a way that the equations become
\begin{equation}
\label{frac2espc}
i \frac{\partial}{\partial t} \psi_1(x,t) = -\frac{1}{2} \frac{\partial^{\mu}}{\partial |x|^{\mu}} \psi_1(x,t) + \gamma e^{i \omega t} \psi_2(x,t),
\end{equation}
 and
 \begin{equation}
i \frac{\partial}{\partial t}  \psi_2(x,t) = -\frac{1}{2} \frac{\partial^{\mu}}{\partial |x|^{\mu}} \psi_2(x,t) + \gamma e^{-i \omega t} \psi_1(x,t).
\label{frac2espac}
\end{equation}
This extension for the set of Schr\"odinger equation essentially 
considers
$\partial_{x}^{2}(\cdots)\rightarrow \partial_{|x|}^{\mu}(\cdots)$ with $1<\mu<2$, 
which corresponds to a Riesz-Weyl fractional operator. By applying the Fourier 
transform in the previous set of equations and using the property 
${\mathcal{F}}\left\{\partial_{|x|}^{\mu}\psi_{1,2}(x,t);k\right\}=-|k|^{\mu}\widetilde{\psi}_{1,2}(k,t)$, we have that
\begin{equation}
 \label{standard2aa1}
i \frac{\partial}{\partial t} \widetilde{\psi}_1(k,t) = \frac{1}{2}|k|^{\mu}  \widetilde{\psi}_1(k,t) + \gamma e^{i \omega t} \widetilde{\psi}_2(k,t),   
\end{equation}
and
 \begin{equation}
i \frac{\partial}{\partial t}  \widetilde{\psi}_2(k,t) = \frac{1}{2} |k|^{\mu} \widetilde{\psi}_2(k,t) + \gamma e^{-i \omega t} \widetilde{\psi}_1(k,t)\;.
\label{standard2bb1}
\end{equation}

By performing some calculations, it is possible to show that
\begin{eqnarray}
\label{F111}
i \frac{\partial}{\partial t} \widetilde{\psi}_1(k,t) = \frac{1}{2}|k|^{\mu}  \widetilde{\psi}_1(k,t) - i\gamma^{2}\int_{0}^{t}dt'e^{-\frac{1}{2}i|k|^{\mu}(t-t')}e^{i\omega (t-t')}\widetilde{\psi}_{1}(k,t')\;,
\end{eqnarray}
which can be solved by using the Laplace transform.

The wave functions for this case can be obtained and are written as
\begin{eqnarray}
\widetilde{\psi}_2(k,t)=-i\gamma\int_{0}^{t}dt'e^{-\frac{1}{2}i|k|^{\mu}(t-t')}e^{-i\omega t'}\widetilde{\psi}_{1}(k,t')\;,
\label{standard2bbb1}
\end{eqnarray}
and 
\begin{eqnarray}
\widetilde{\psi}_{1}(k,t)=e^{-\frac{1}{2}i\left(|k|^{\mu}-\omega\right)t}\left[\cos\left(\frac{1}{2}t\sqrt{\omega^{2}+4\gamma^{2}}\right)-\frac{i\omega}{\sqrt{\omega^{2}+4\gamma^{2}}}\sin\left(\frac{1}{2}t\sqrt{\omega^{2}+4\gamma^{2}}\right)\right]\widetilde{\varphi}_{1}(k)\;,
\label{standard2aaaa1}
\end{eqnarray}
assuming the initial conditions $\widetilde{\psi}_{1}(k,0)=\widetilde{\varphi}_{1}(k)$ and  $\widetilde{\psi}_{2}(k,0)=0$. The inverse Fourier transform of Eqs.~(\ref{standard2bbb1}) and~(\ref{standard2aaaa1}) results in 
\begin{eqnarray}
\psi_{1}(x,t)=\Xi_{1}(t)\int_{-\infty}^{\infty}dx'{\cal{G}}_{\mu}(x-x',t)\varphi_{1}(x'),
\label{standard2aaaaa}
\end{eqnarray}
and
\begin{eqnarray}
\psi_{2}(x,t)=\Xi_{2}(t)\int_{-\infty}^{\infty}dx'{\cal{G}}_{\mu}(x-x',t)\varphi_{1}(x'),
\label{standard2bbbbb}
\end{eqnarray}
with
\begin{equation}
{\cal{G}}_{\mu}(x,t)= \frac{1}{|x|}{\mbox{H}}^{1,1}_{2,2}
\left[\frac{2}{it}|x|^{\mu}
\left|^{\left(1, 1\right), \left(1, \frac{\mu}{2}
\right)}_{\left(1, \mu\right), \left(1, \frac{\mu}{2}
\right)}\right.\right],
\label{GLevy_Chap8}
\end{equation}
which resembles the form of the L\'evy distribution found in anomalous 
diffusion processes. In Eq.~(\ref{GLevy_Chap8}), we have 
the H Fox function~\cite{Fox}, usually represented~\cite{Book01} by  
\begin{eqnarray}
\mbox {\large{H}}^{m,n}_{p,q}
\bigg[ z \bigg|
\begin{array}{c}
\left(a_{p},A_{p}\right)\\
\left(b_{q},B_{q}\right)\\
\end{array}
\bigg]&=&\mbox {\large{H}}^{m,n}_{p,q}
\bigg[ z \bigg|
\begin{array}{c}
\left(a_{1},A_{1}\right)\cdots\left(a_{p},A_{p}\right)\\
\left(b_{1},B_{1}\right)\cdots\left(b_{q},B_{q}\right)\\
\end{array}
\bigg]=\frac{1}{2\pi i}\int_{L}ds \chi(s) z^{s}
\end{eqnarray} 
where
\begin{equation}
\chi(s)=\frac{\prod_{j=1}^{m}\Gamma\left(b_{j}-B_{j}s\right)\prod_{j=1}^{n}\Gamma\left(1-a_{j}+A_{j}s\right)}{\prod_{j=1}^{q}\Gamma\left(1-b_{j}+B_{j}s\right)\prod_{j=1}^{p}\Gamma\left(a_{j}-A_{j}s\right)}\;,    
\end{equation}

which involves Mellin--Barnes integrals~\cite{Book01}.
The asymptotic behavior of Eq.~(\ref{GLevy_Chap8}) in the limit of 
$|x|\rightarrow \infty$ is given by
${\cal{G}}_{\mu}(x,t)\sim i \left[t/\left(2 |x|^{1+\mu}\right)\right]$, 
which is different from the usual one characterized by the Gaussian 
behavior. Note that this result for the asymptotic limit 
can be obtained by using the approach employed in Ref.~\cite{saxena2006fractional}. 
It is essentially an integration over the Mellin - Branes integral poles, 
which represents Eq.~(\ref{GLevy_Chap8}). This feature is directly connected 
to the presence of spatial fractional differential operators in Eqs.~(\ref{standard2bb1}) and~(\ref{standard2bbb1}).

In addition to the analytical approach, it is possible to investigate 
the dynamical behavior from the numerical point of view by using the 
following discretization
\begin{equation}
\psi^{i,j+1}_1 = \psi^{i,j}_1 + i\xi_\mu\sum_{k=0}^{i-1}[\psi^{i-k+1,j}_1 - 2\psi^{i-k,j}_1 + \psi^{i-k-1,j}_1][(k+1)^{2-\mu} - k^{2-\mu}] - i\beta(V^{j}_1 \psi^{i,j}_2 + V^{j+1}_1 \psi^{i,j+1}_2),
\end{equation}
and
\begin{equation}
 \psi^{i,j+1}_2 = \psi^{i,j}_2 + i\xi_\mu\sum_{k=0}^{i-1}[\psi^{i-k+1,j}_2 - 2\psi^{i-k,j}_2 + \psi^{i-k-1,j}_2][(k+1)^{2-\mu} - k^{2-\mu}] - i\beta(V^{j}_2 \psi^{i,j}_1 + V^{j+1}_2 \psi^{i,j+1}_1),    
\end{equation}

where $\xi_\mu \equiv \Delta t /2\Gamma(3-\mu)\Delta x^{\mu}$, $\beta \equiv \gamma\Delta t / 2$, $V_1 = e^{i\omega t}$ and $V_2 = e^{-i \omega t}$ \cite{shen2005}.

\begin{figure}[hbt]
\centering
\includegraphics[scale=0.28]{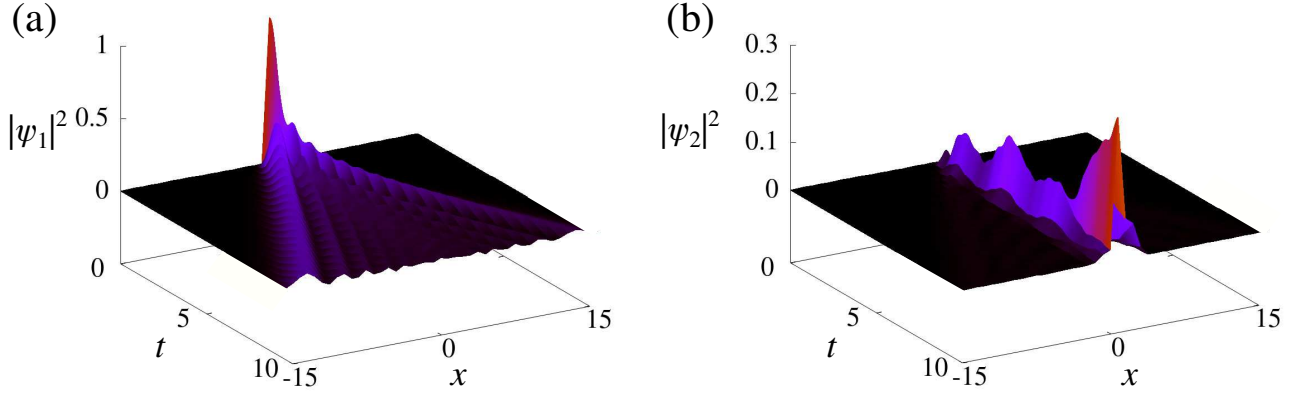}
\caption{Probability distribution with a fractional kinetic term of finding 
the system in $\psi_1$ state, in the panel (a), and in $\psi_2$ state, in 
the panel (b). We consider $\mu = 1.95$, $\xi_\mu = 10^{-5}$, $\beta = 10^{-5}$, 
$\gamma=0.5$, $\sigma=0.4$, $\omega=2\pi$, $\Delta x = 1.0$, $\Delta t = 0.000125$.}
\label{fig9}
\end{figure}

Small changes in the order of the fractional space operator 
make significant changes in the spread probability dynamics. This phenomenon 
is observed in Fig.~\ref{fig9}, which exhibits the spread of $|\psi_1|^2$ 
in the panel (a) and $|\psi_2|^2$ in the panel (b). The Gaussian package 
is the initial condition spread widely in the presence of a space fractional 
operator compared with the standard case. Another notable characteristic 
is the behavior of $|\psi_2|^2$. The probability of finding the system in 
$\psi_2$ state is more centered in origin and has a higher probability 
when compared with the previous cases. The $\psi_2$ state assumes the 
Gaussian shape probability and, for long times, replicates the dynamics 
observed in $\psi_1$.

\begin{figure}[hbt]
\centering
\includegraphics[scale=0.28]{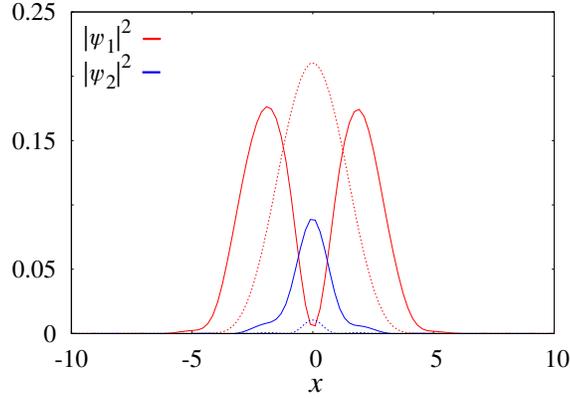}
\caption{Probability distribution at $t = 1.5$ for the states $\psi_1$ 
(red line) and $\psi_2$ (blue line). The continuous line shows the 
behavior for $\mu = 1.95$ and the dotted for $\mu = 2$. We consider 
$\mu = 1.95$, $\xi_\mu = 10^{-5}$, $\beta = 10^{-5}$, $\gamma=0.5$, 
$\sigma=0.4$, $\omega=2\pi$, $\Delta x = 0.25$, $\Delta t = 0.000125$.}
\label{fig10}
\end{figure}

The comparison of the probabilities in $t = 1.5$ is shown in Fig.~\ref{fig10}, 
where the continuous and dotted lines represent the cases $\mu = 1.95$ and 
$\mu = 2.0$, respectively. The result shows a sharper division in the 
Gaussian package along with an enlargement in the package. The probability 
of the $\psi_2$ is more centered in origin, indicating that this state 
will take over a Gaussian behavior over time.

\begin{figure}[hbt]
\centering
\includegraphics[scale=0.28]{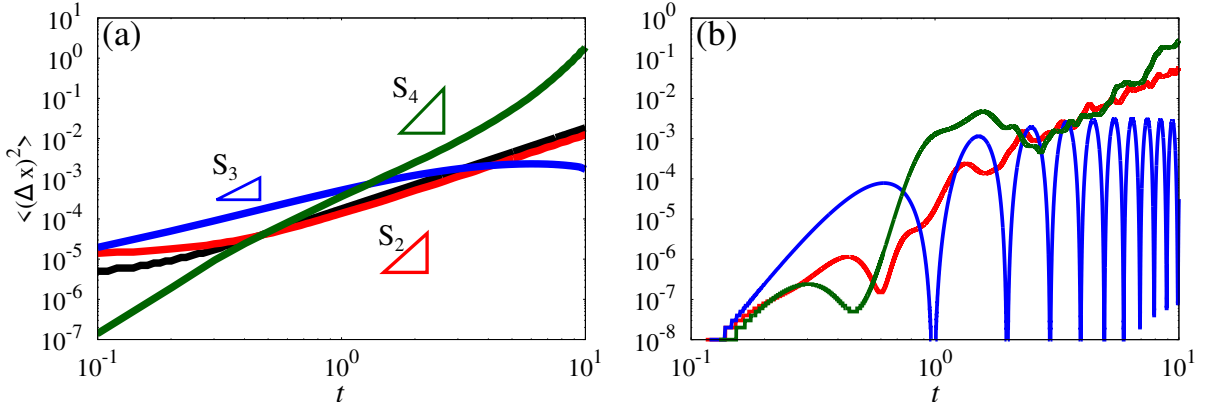}
\caption{Mean square displacement for the Gaussian package. Panel (a) 
is for $\psi_1$ state and panel (b) is for $\psi_2$ state. The green 
points are for $\mu = 1.95$, the red for the standard case ($S_2 = 2.07$), 
the blue for $\alpha = 0.98$ ($S_3 = 1.87$), and the black for free-particle. 
The slope associated with the green curve is $S_4=2.61$. We consider 
$\mu = 1.95$, $\xi_\mu = 10^{-4}$, $\beta = 10^{-5}$, $\gamma=0.5$, 
$\sigma=0.4$, $\omega=2\pi$, $\Delta x = 0.66$, $\Delta t = 0.000125$.}
\label{fig11}
\end{figure}

The mean square displacement for the  fractional derivative in space is 
shown in Fig.~\ref{fig11}, with the green points in Fig.~\ref{fig11}(a) 
for $\psi_1$ and by the green line in Fig.~\ref{fig11}(b) for $\psi_2$. 
The red points are for the fractional time derivative ($\alpha = 0.98$), 
the blue points for the fractional space derivative ($\mu = 1.95$),  and 
the black points for the free-particle. The $|\psi_1|^2$ spread 
as $\sim t^{S_4}$ with $S_4 = 2.61$. Compared with other 
cases, the fractional space operator makes the probability package spread 
more widely, i.e., if we consider $\mu=2$ in a determined time, the package 
occupies a certain range of space; however, for the same situation with 
$\mu=1.95$ the package occupies a larger range. As shown in 
Fig.~\ref{fig11}(b), the $|\psi_2|^2$ spread  more intensely. Populating 
the $\psi_2$ state follows the Gaussian shape, as noted in Fig.~\ref{fig9}(b).

The last possible case to be analyzed is the Schr\"odinger equation with 
fractional differential operators in space and time by taking into 
account a time-dependent potential, i.e.,
\begin{equation}
 \label{frac2aa}
i^{\alpha} \frac{\partial^{\alpha}}{\partial t^{\alpha}} \psi_1(x,t) = -\frac{1}{2} \frac{\partial^{\mu}}{\partial |x|^{\mu}} \psi_1(x,t) + \gamma e^{i \omega t} \psi_2(x,t),   
\end{equation}
and
\begin{equation}
i^{\alpha} \frac{\partial^{\alpha}}{\partial t^{\alpha}}  \psi_2(x,t) = -\frac{1}{2} \frac{\partial^{\mu}}{\partial |x|^{\mu}} \psi_2(x,t) + \gamma e^{-i \omega t} \psi_1(x,t)\;,
\label{frac2bb}
\end{equation}
It is possible to find a solution for these equations, and it is given by
\begin{eqnarray}
\label{sol2a}
\psi_{1}(x,t)&=&\psi_{1}^{(0)}(x,t)+
\sum_{n=1}^{\infty}\left(\gamma/ i^{\alpha}\right)^{2n}\int_{-\infty}^{\infty}dx_{n}\int_{0}^{t}dt_{n}
{\Upsilon}(x-x_{n},t-t_{n})\nonumber \\ &\times&\int_{-\infty}^{\infty}dx_{n-1}\int_{0}^{t_{n}}dt_{n-1}{\Upsilon}(x_{n}-x_{n-1},t_{n}-t_{n-1})\cdots
\int_{-\infty}^{\infty}dx_{1}\int_{0}^{t_{2
}}dt_{1}{\Upsilon}(x_{2}-x_{1},t_{2}-t_{1})\psi_{1}^{(0)}(x_{1},t_{1})
\end{eqnarray}
with
\begin{eqnarray}
\label{sol2b}
\psi_{2}(x,t)=\frac{\gamma}{i^{\alpha}}\int_{-\infty}^{\infty}dx'\int_{0}^{t}dt' t'^{\alpha-1}{\cal{G}}_{\alpha,\mu}^{(\alpha)}(x-x',t-t')e^{i\omega t'}\psi_1(x',t')\;,
\end{eqnarray}
where $\psi_{1}^{(0)}(x,t)=\int_{-\infty}^{\infty}dx'\varphi(x'){\cal{G}}_{\alpha,\mu}^{(1)}(x-x',t)$, 
\begin{eqnarray}
\Upsilon(x,t)=e^{i\omega t}\int_{-\infty}^{\infty}dx'\int_{0}^{t}dt't'^{\alpha-1}e^{-i\omega t'}{\cal{G}}_{\alpha,\mu}^{(\alpha)}(x',t'){\cal{G}}_{\alpha,\mu}^{(\alpha)}(x-x',t-t')\;,
\end{eqnarray}
and
\begin{eqnarray}
{\cal{G}}_{\alpha,\mu}^{(\beta)}(x,t)=\frac{1}{|x|}{\mbox{H}}^{2,1}_{2,3}
\left[-\frac{|x|^{\mu}}{t^{\alpha}/(2i^{\alpha})}
\left|^{\left(1, 1\right),\left(\beta,\alpha\right), \left(1, \frac{\mu}{2}
\right)}_{\left(1,\mu\right),\left(1, 1\right), \left(1, \frac{\mu}{2}
\right)}\right.\right]\;.
\label{Green1}
\end{eqnarray}
Note that Eq.~(\ref{Green1}) is essentially the Green function of this 
case and, consequently, connected to the relaxation process of this system. 
It differs from the previous case since it mixes different fractional operators in space and time. 

\begin{figure}[hbt]
\centering
\includegraphics[scale=0.28]{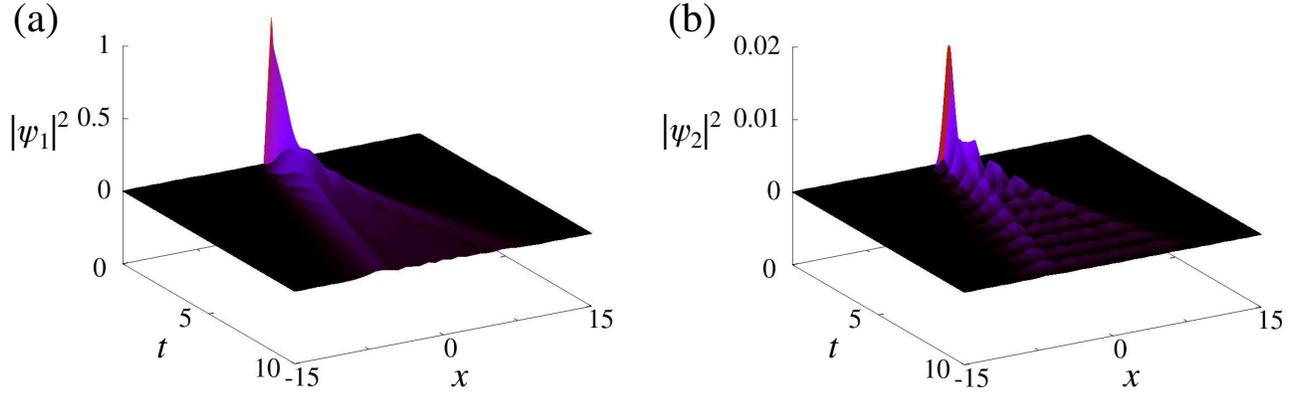}
\caption{Probability distribution with a fractional kinetic term of 
finding the system in $\psi_1$ state, in the panel (a), and in $\psi_2$ 
state, in the panel (b) for time and space fractional dependence. We 
consider $\alpha = 0.98$, $\mu = 1.95$, $\gamma=0.5$, $\sigma=0.4$, 
$\omega=2\pi$, $\xi_{\alpha,\mu} = 6 \times 10^{-4}$, 
$\beta_{\alpha,\mu} = \times 10^{-4}$, $\Delta x = 0.66$, $\Delta t = 5 \times 10^{-4}$.}
\label{fig12}
\end{figure}

It is possible to write a combination of the previous discretizations 
schemes and find the equations
\begin{eqnarray}
\psi^{i,j+1}_1 &=& \psi^{i,j}_1 - \sum_{k=1}^{j} [(k+1)^{(1-\alpha)} - k^{(1-\alpha)}][\psi^{i,j+1-k}_{1} - \psi^{i,j-k}_1] + i^{-\alpha}\beta_{\alpha,\mu}(V^{j}_1 \psi^{i,j}_2 + V^{j+1}_1 \psi^{i,j+1}_2) \nonumber \\ &-& i^{-\alpha}\xi_{\alpha,\mu}\sum_{k=0}^{i-1}[\psi^{i-k+1,j}_1 - 2\psi^{i-k,j}_1 + \psi^{i-k-1,j}_1][(k+1)^{2-\mu} - k^{2-\mu}],
\end{eqnarray}
and
\begin{eqnarray}
\psi^{i,j+1}_2 &=& \psi^{i,j}_2 - \sum_{k=1}^{j} [(k+1)^{(1-\alpha)} - k^{(1-\alpha)}][\psi^{i,j+1-k}_{2} - \psi^{i,j-k}_2] + i^{-\alpha}\beta_{\alpha,\mu}(V^{j}_2 \psi^{i,j}_1 + V^{j+1}_2 \psi^{i,j+1}_1) \nonumber \\ &-& i^{-\alpha}\xi_{\alpha,\mu}\sum_{k=0}^{i-1}[\psi^{i-k+1,j}_2 - 2\psi^{i-k,j}_2 + \psi^{i-k-1,j}_2][(k+1)^{2-\mu} - k^{2-\mu}],
\end{eqnarray}
where $\xi_{\alpha,\mu} = \Gamma(2-\alpha) \Delta t^{\alpha}/2\Gamma(3-\mu)\Delta x^\mu$, $\beta_{\alpha,\mu} = \gamma \Delta t^{\alpha} \Gamma(2-\alpha)/2$, $V_1 = e^{i\omega t}$ and $V_2 = e^{-i \omega t}$. Considering $\alpha = 0.98$ and $\mu = 1.95$ the results for probability distribution is shown in Fig.~\ref{fig12}(a) for $|\psi_1|^2$ and in \ref{fig12}(b) for $|\psi_2|^2$. The results of the combination of both fractional derivatives show a combination of the two previously discussed behavior. However, for this set of parameters, the results resemble fractional time dependence than space one.

\begin{figure}[hbt]
\centering
\includegraphics[scale=0.28]{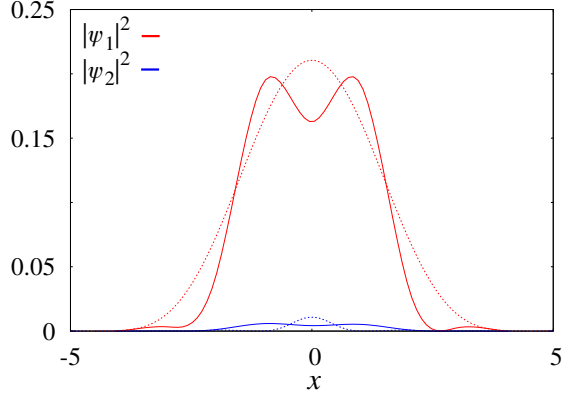}
\caption{Probability distribution at $t = 1.5$ for the states $\psi_1$ 
(red line) and $\psi_2$ (blue line). The continuous line show the behavior 
for $\alpha = 0.98$, $\mu = 1.95$ and the dotted for $\alpha = 1$ and $\mu = 2$.    
We consider $\alpha = 0.98$, $\mu = 1.95$, $\gamma=0.5$, $\sigma=0.4$, 
$\omega=2\pi$, $\xi_{\alpha,\mu} = 6 \times 10^{-4}$, 
$\beta_{\alpha,\mu} = \times 10^{-4}$, $\Delta x = 0.66$, $\Delta t = 5 \times 10^{-4}$.}
\label{fig13}
\end{figure}

Figure~\ref{fig13} displays the comparison between the probability 
distribution for fractional time and space order (in continuous lines) 
versus the standard model. These results make it possible 
to verify the composition of both fractional operators. To gain greater 
influence for $\mu$ or $\alpha$ in the dynamics, it is necessary to 
decrease one of these values. 

\begin{figure}[hbt]
\centering
\includegraphics[scale=0.28]{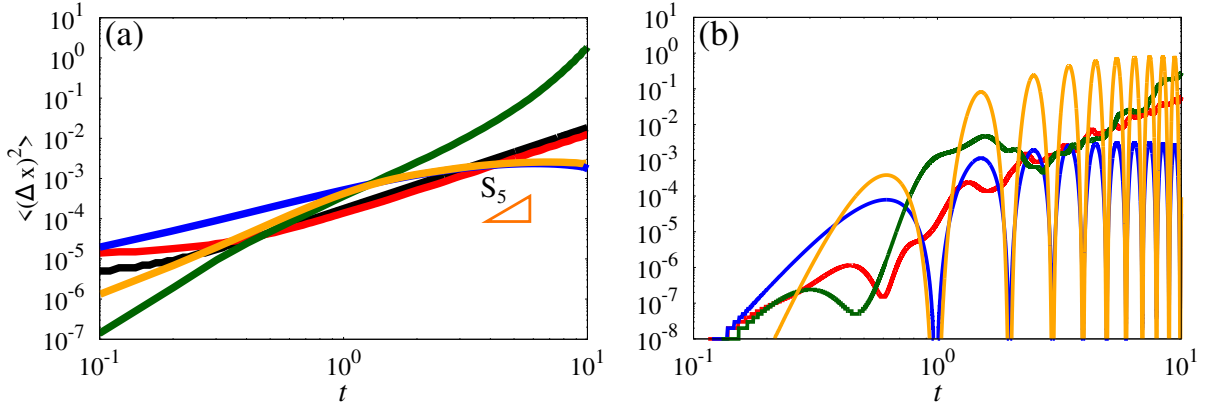}
\caption{Mean square displacement for the Gaussian package. Panel (a) is 
for $\psi_1$ state, and panel (b) is for $\psi_2$. The orange points are 
for $\alpha = 0.98$ and $\mu = 1.95$, the green points for $\mu = 1.95$, 
the red for $\alpha = 0.98$, and the black for a free particle. The 
slope associated with $\alpha = 0.98$ and $\mu = 1.95$ is $S_5 = 1.76$. 
We consider $\alpha = 0.98$, $\mu = 1.95$, $\gamma=0.5$, $\sigma=0.4$, 
$\omega=2\pi$, $\xi_{\alpha,\mu} = 6 \times 10^{-4}$, 
$\beta_{\alpha,\mu} = \times 10^{-4}$, $\Delta x = 0.66$, $\Delta t = 5 \times 10^{-4}$.}
\label{fig14}
\end{figure}

The spreading of the Gaussian package for $\alpha < 1$ and $\mu < 2$ is 
more centered than in the standard case, as we see in 
Fig.~\ref{fig14}(a) -- orange points. Figure~\ref{fig14}(b) displays 
the spread for $\psi_2$ state by the orange line. The associated 
slope is equal to $S_5 = 1.76$, near $S_3 = 1.87$, obtained for the case when 
only the time fractional operator was considered. Therefore, in the presence 
of both fractional operators, the time derivative supplants the effects 
of the space derivative. The orange and blue points match in the range 
displayed in Fig.~\ref{fig14}(a) after some $t$. However, the distribution 
for $|\psi_2|^2$, corresponding to the orange line in the panel \ref{fig14}(b), 
follows the same shape as in the case of the fractional time derivative.


\section{Conclusion}
We analyzed the influence of fractional operators in the Schr\"odinger 
equation when an oscillating time-dependent potential is considered to 
simulate an oscillatory external field applied in the system. We 
started with a two-level system, which was first analyzed by considering 
the static case $\omega=0$ and after the time-dependent case $\omega\neq 0$. 
We obtained time analytical and numerical solutions for the standard and 
the fractional cases. In particular, we verified that the solutions had 
an oscillating behavior for a long time. Afterward, we incorporated the 
kinetic term in the Hamiltonian to allow the spreading of the system. We 
also considered one state populated as an initial condition while the 
other remained empty. We also analyzed this scenario from the analytical 
and numerical points of view for the standard and the fractional cases. 
For the fractional cases, we first consider the effect of the fractional 
time derivatives, and after analyzing
the spatial fractional derivatives, which preserve the probability of 
the system. One of them is the non-conservation of the probability of 
the system. We analyzed the behavior of the mean square displacement 
(or deviation) for these cases and compared it with the free particle case. 
The results showed that the fractional differential operators lead to 
different behavior for spreading the system when compared with the 
standard case. For the fractional derivative in space, we have 
a faster spreading of the initial condition. On the other hand, we see 
a slower spreading of the wave package when fractional derivatives in 
time are incorporated in the Schrodinger equation. This feature is 
also present in the diffusion context when fractional differential 
operators are considered, evidencing that these operators strongly influence 
the random process connected to these phenomena. The mean square displacement 
also evidenced this point and the influence on the uncertain relations, 
as observed by Laskin~\cite{PhysRevE.66.056108}.

\section*{Acknowledgements}

 The authors thank the financial support from the Brazilian Federal 
 Agencies (CNPq), the S\~ao Paulo Research
Foundation (FAPESP, Brazil), CAPES, Funda\-\c c\~ao A\-rauc\'aria.
The authors thank the 105 Group Science (www.105groupscience.com). E.K.L. acknowledges the support of the CNPq (Grant No. 301715/2022-0).

\section*{Appendix I}

A numerical method to solve initial-problem based on Caputo definition is 
a generalization of the classical Adams-Bashforth-Moulton. This method 
was proposed by Diethelm, Ford and Freed \cite{diethelm2005}, and is 
defined by the follows equations:
\begin{equation}
    y_h (t_{n+1}) = \sum_{k=0}^{\left \lceil{\alpha}\right \rceil } \frac{t^k_{n+1}}{k!} y_0^{(k)} + \frac{h^\alpha}{\Gamma(\alpha + 2)} f(t_{n+1},y^P_h (t_{n+1})) + \frac{h^\alpha}{\Gamma(\alpha + 2)} \sum_{j=0}^n a_{j,n+1} f(t_j , y_h (t_j)),
\end{equation}
where
\begin{equation}
    y^P_{h} (t_{n+1}) = \sum_{k=0}^{\left \lceil{\alpha}\right \rceil } \frac{t^k_{n+1}}{k!} y_0^{(k)} + \frac{1}{\Gamma(\alpha)} \sum_{j=0}^n b_{j,n+1} f(t_j, y_h (t_j)).
\end{equation}
The coefficients are defined by
\begin{equation}
    a_{j,n+1} =
    \left\{\begin{array}{lll}
    n^{\alpha + 1} - (n - \alpha)(n+1)^\alpha, \ \ {\rm if} \ j = 0, \\
    (n-j+2)^{\alpha+1} + (n-j)^{\alpha + 1} - 2(n-j+1)^{\alpha+1}, \ \ {\rm if}  \ 1 \leq j \leq n\\
    1, \ \ {\rm if} j = n + 1,
    \end{array}\right.
\end{equation}
and
\begin{equation}
    b_{j,n+1} = \frac{h^\alpha}{\alpha}((n+1-j)^{\alpha} - (n-j)^\alpha),
\end{equation}
where $j=1,2,...,n$ and $n$ is associated with discrete time window, $T$, 
which is discrete in $t_n = nh$, with $n = 0,1,...,N$, where $T = Nh$.


\end{document}